\documentclass[prc,twocolumn,showpacs,floatfix,nofootinbib,preprintnumbers,superscriptaddress,amsmath,amssymb]{revtex4-1}

\usepackage{graphicx}
\usepackage{epsfig}
\usepackage{amsmath,amssymb}
\usepackage{bm}
\usepackage{color}
\usepackage{float}
\usepackage{dcolumn}

\graphicspath{ {images/} }
\newcommand{\be}{\begin{equation}}
\newcommand{\ee}{\end{equation}}
\newcommand{\ba}{\begin{array}}
\newcommand{\ea}{\end{array}}
\newcommand{\bn}{\begin{eqnarray}}
\newcommand{\en}{\end{eqnarray}}
\newcommand{\bnl}{\begin{mathletters}\begin{eqnarray}}
\newcommand{\enl}{\end{eqnarray}\end{mathletters}}
\newcommand{\bml}{\begin{mathletters}}
\newcommand{\eml}{\end{mathletters}}
\newcommand{\bc}{\begin{center}}
\newcommand{\ec}{\end{center}}
\newcommand{\bi}{\begin{itemize}}
\newcommand{\ei}{\end{itemize}}
\newcommand{\bt}{\begin{tabular}}
\newcommand{\et}{\end{tabular}}
\newcommand{\bnll}[1]{\begin{subequations}\label{#1}\begin{eqnarray}}
\newcommand{\enll}{\end{eqnarray}\end{subequations}}

\newcommand{\thalf}{\tfrac{1}{2}}

\renewcommand{\bbox}[1]{\bm{#1}}
\newcommand{\mathsfI}{\delta}

\newcommand{\particle}{\mbox{p-h}}

\def\bra{\langle\,}
\def\ket{\,\rangle}

\begin{document}

\title{Isospin-invariant Skyrme energy-density-functional approach with axial symmetry}
\date{\today}

\author{J.A. Sheikh}
\affiliation{Department of Physics and
  Astronomy, University of Tennessee, Knoxville, Tennessee 37996-1200, USA}
\affiliation{Joint Institute for Heavy-Ion Research, Oak Ridge, Tennessee 37831-6374, USA}
\affiliation{Department of Physics, University of Kashmir,
Srinagar, 190 006, India}

\author{N. Hinohara}
\affiliation{Department of Physics and Astronomy,
University of North Carolina, Chapel Hill, North Carolina 27599-3255, USA}
\affiliation{Department of Physics and
  Astronomy, University of Tennessee, Knoxville, Tennessee 37996-1200, USA}
\affiliation{Joint Institute for Heavy-Ion Research, Oak Ridge, Tennessee 37831-6374, USA}

\author{J. Dobaczewski}
\affiliation{Institute of Theoretical Physics, Faculty of Physics, University of Warsaw,
Ho{\.z}a 69, 00-681 Warsaw, Poland}
\affiliation{
Department of Physics, PO Box 35 (YFL)
University of Jyv{\"a}skyl{\"a}, FI-40014 Jyv{\"a}skyl{\"a}, Finland}
\affiliation{Department of Physics and
  Astronomy, University of Tennessee, Knoxville, Tennessee 37996-1200, USA}

\author{T. Nakatsukasa}
\affiliation{RIKEN Nishina Center, Wako 351-0198, Japan}
\affiliation{Center for Computational Sciences, University of Tsukuba,
Tsukuba 305-8571, Japan}

\author{W. Nazarewicz}
\affiliation{Department of Physics and
  Astronomy, University of Tennessee, Knoxville, Tennessee 37996-1200, USA}
\affiliation{Physics Division, Oak Ridge National Laboratory, P.O. Box
  2008, Oak Ridge, Tennessee 37831-6373, USA}
\affiliation{Institute of Theoretical Physics, Faculty of Physics, University of Warsaw,
Ho{\.z}a 69, 00-681 Warsaw, Poland}

\author{K. Sato}
 \affiliation{RIKEN Nishina Center, Wako 351-0198, Japan}

\begin{abstract}
\begin{description}

\item[Background]
Density functional theory (DFT) is the
microscopic tool of choice to describe  properties of nuclei over the entire
nuclear landscape, with a focus on medium-mass and heavy complex systems.
Modern energy density functionals (EDFs) often offer  a level of accuracy
typical of phenomenological approaches based on parameters locally
fitted to the data. It is clear, however, that in order to achieve high quality
of predictions to guide spectroscopic studies, current functionals must be improved, especially  in the isospin channel. In this respect, experimental studies of short-lived  nuclei far from
stability offer a unique test of isospin aspects of the many-body theory.

\item[Purpose]
We  develop the
isospin-invariant Skyrme-EDF method by considering local densities in all possible isospin channels and proton-neutron (p-n)  mixing terms as mandated by the isospin symmetry. The  EDF employed
has the most general form that
depends quadratically on the isoscalar
and isovector densities. We test and benchmark the resulting p-n EDF approach, and study the general properties of the new scheme by means of the cranking in the isospin space.

\item[Methods]
We extend the existing axial DFT solver {\sc hfbtho} to the case of isospin-invariant
EDF approach with all possible p-n mixing terms.
Explicit expressions
have been derived for all the densities and potentials that appear in the
isospin representation. In practical tests, we consider the Skyrme EDF SkM$^*$ and, as a first
application,
concentrate on Hartree-Fock aspects of the problem, i.e.,  pairing has been disregarded.

\item[Results]
Calculations have been performed for the ($A=78, T\simeq 11$),
($A = 40, T\simeq 8$), and ($A = 48, T\simeq 4$)  isobaric analog chains. Isospin structure of self-consistent p-n mixed solutions
has been investigated with and without the Coulomb
interaction, which is the sole source of isospin symmetry breaking in our approach. The extended axial  {\sc hfbtho} solver has been benchmarked against  the symmetry-unrestricted {\sc hfodd} code for deformed and spherical states.

\item[Conclusions]
We developed and tested a general isospin-invariant Skyrme-EDF framework. The new approach permits spin-isospin densities that may give rise to, hitherto, unexplored modes in the excitation spectrum. The new formalism has been tested in the Hartree-Fock limit.
A systematic comparison between {\sc hfodd} and {\sc hfbtho} results show a maximum deviation of about
10\,keV on the total binding energy for deformed nuclei when the Coulomb term is included. Without this term, the results of both solvers agree down to a $\sim$10\,eV level.
\end{description}

\end{abstract}

\pacs{21.10.Hw,21.60.Jz,21.10.Sf}

\maketitle

\section{INTRODUCTION \label{sec:intro}}

A major challenge for low-energy nuclear theory is to develop a universal nuclear EDF that can be used to explain and predict static and dynamic properties of atomic nuclei throughout the entire nuclear landscape within the framework of nuclear DFT. In a worldwide effort to develop a  general-purpose nuclear EDF \cite{(Sto10),(Dob12),(Bog13)}, various strategies are applied, and, to realize this vision, the properties of rare isotopes are an essential guide.

In the quest of developing a universal nuclear EDF,
the existing  functionals ought to be enriched by
incorporating the neglected couplings, especially in the spin
and isospin channels. Indeed, the recent work \cite{(Kor13)}
suggests that the Skyrme EDF has reached its limits and significant changes to the form of the functional are needed.
As far as the isospin sector is concerned,
 most of the EDFs include isoscalar particle densities and a single $t_z$ component in the isovector channel. The  $t_x$ and
$t_y$, or p-n mixed, components of isovector densities are
completely neglected. In the heavier nuclei where neutrons and
protons occupy different shell-model spaces, the neglect of the
p-n mixed densities could be  justified. However, in the
lighter and medium-mass nuclei, neutrons and protons
move in the same shells and the exclusion of these isovector densities
cannot be justified. There are several observations that indicate
deficiencies inherent in the existing EDF and other
approaches to describe nuclei in the vicinity of the $N = Z$ line \cite{(Sat06),(Afa13a)}.
For instance, it is quite well established that binding energies of
the nuclei close to the $N = Z$ line are underestimated  by theoretical
models~\cite{(Zel96),(Sat97)}, and p-n correlations are expected to be the
missing piece of physics in this puzzle.

In some earlier studies, p-n mixing has been investigated
in the particle-particle channel~\cite{(Goo79),(Ter98),(Fra99),(She00),(Sat01),(Gez11)} (see Refs. \cite{(Sat06),(Afa13a)} and Sec. VI of Ref.~\cite{(Roh10)} for a more complete list of references). In the
particle-hole sector, however, the p-n mixing and resulting symmetry breaking effects have  been largely neglected (a notable exception is the recent study of $^{21}$Ne~\cite{(Rob11)} that considered a p-n mixing on the HF level). As discussed in \cite{(Roh10)}, such an approximation does
not seem to be justified as the self-consistent polarization
between particle-hole (p-h) and particle-particle (p-p) HFB channels is  known to be strong.
In Refs.~\cite{(Per04),(Roh10)}
a generalized EDF approach has been proposed
that allows for the arbitrary mixing of protons and neutrons, and an  isospin-invariant
EDF has been constructed. It has been
shown that the generalized EDF gives rise to novel spin-isospin
combinations of nucleonic  densities that are absent in the standard Skyrme
approaches. We expect that those extensions  may lead to new, hitherto unexplored, nuclear modes.

The main objective of this study  is to develop, test, and benchmark  the isospin-invariant
Skyrme-EDF  (pnEDF) approach formulated in the cylindrical coordinate
system, whose building blocks are all
possible p-n mixed local densities.
Since the majority of nuclei are axial in their ground states,
such an approach will allow us to extend the global surveys of nuclear properties \cite{(Sto03),(Erl12),(Kor13a),(Ols13),*(Ols13a)} made with the axial DFT solver {\sc hfbtho}~\cite{(Sto05),(Sto13)} to observables and decays  related to isospin.
The code {\sc hfbtho} has been optimized for performance on flagship computing platforms, for it serves as a backbone of the EDF optimization package \cite{(Kor10),*(Kor12),(Kor13)}.
In a parallel
study~\cite{(Sat13c)}, p-n mixed densities have also been
implemented in the general-purpose  solver {\sc hfodd}~\cite{(Sch12),*(Sch14)}  written in a three-dimensional
Cartesian basis. We take advantage of this development to benchmark both
pnEDF schemes.

The paper is organized as follows. Basic
expressions pertaining to the isospin-invariant  pnEDF approach are
briefly summarized  in Sec.~\ref{sec:formula}. Section~\ref{sec:model}
discusses the HF application of the formalism to isobaric analog states (IASs)
using  the two-dimensional isocranking formalism. The  axial
 {\sc hfbtho} pnEDF framework is benchmarked against the symmetry-unconstrained
pnEDF   {\sc hfodd}  approach in Sec.~\ref{sec:comparison}.
Finally, Sec.~\ref{sec:summary} contains the summary of our
work and prospects for  further developments.

\section{Basics of \lowercase{pn}EDF approach}\label{sec:formula}
The pnEDF Kohn-Sham state $|\Psi\rangle$ is a single Slater determinant
built of the set of $A$ fully occupied single-particle (s.p.) states, that is,
\begin{align}
|\Psi\rangle = \prod_{k=1}^A c^{+}_{k}|0\rangle,
\end{align}
where $c^{+}_k$ denotes the s.p.\ state creation operator. This operator
can be expressed in terms of the s.p.\ wave function $V_k(\bbox{r}st)$,
\begin{align}
  c_k^{+} = \int d^3 \bbox{r} \sum_{st} V^\ast_k(\bbox{r}st) a^{+}_{\bbox{r}st} \quad  (k \le A),
\end{align}
where $a^{+}_{\bbox{r}st}$ creates the nucleon at point $\bbox{r}$,
spin $s$=$\pm\thalf$, and isospin $t=+\frac{1}{2}$ (neutron)  or
$-\frac{1}{2}$ (proton). In the p-n mixing framework, the HF
s.p.\ state $c_k^{+}$ contains the neutron and proton components. In
the present study, we only consider unpaired systems and
particle-hole ({\particle}) densities, whereupon $V_k(\bbox{r}st)$
are simply the  self-consistent HF wave functions. However, expressions
given below are also valid within the HFB approach, where
$V_k(\bbox{r}st)$ correspond to lower components of the quasiparticle
wave functions~\cite{(Rin80),(Dob96)}.

To fix the notation, we now recall basic expressions introduced
and derived in Ref.~\cite{(Per04)}.
The one-body density matrix $\hat{\rho}$ is defined as
\begin{align}
\hat{\rho}(\bbox{r}st,\bbox{r}'s't')
&=\langle \Psi |a_{\bbox{r}'s't'}^{+}a_{\bbox{r}st}|\Psi \rangle \nonumber \\
&= \sum_{k=1}^A V_k(\bbox{r}'s't') V_k^\ast(\bbox{r}st),
\label{eq:rho}
\end{align}
and the pnEDF can be written as
\be
\bar{H}[\hat{\rho}] = \int d^3\bbox{r}\, {\mathcal H}(\bbox{r})
= \int d^3\bbox{r}\, {\mathcal H}_{\rm Sk}(\bbox{r}) + E_{\rm Cou}[\hat{\rho}],
\ee
where the Skyrme energy density is
\be
{\mathcal H}_{\rm Sk}(\bbox{r})
=
\frac{\hbar^{2}}{2m} \tau_0(\bbox{r})
+\chi_0(\bbox{r})+\chi_1(\bbox{r})
\label{enden}
\ee
with $\tau_0(\bbox{r})$ being the isoscalar kinetic-energy density (it is
assumed in the following that the neutron and proton masses are equal). The
Coulomb energy functional $E_{\rm Cou}$ is the only term that breaks
the isospin symmetry.
The Slater approximation is used for the Coulomb exchange functional.
The {\particle} Skyrme interaction-energy densities
$\chi_0(\bbox{r})$ and $\chi_1(\bbox{r})$ depend quadratically on the isoscalar
and isovector densities, respectively. Based on general
rules of constructing the energy density~\cite{(Per04)}, one obtains
\begin{widetext}
\bnll{chiph}
\label{chi0ph}
{\chi}_{0}(\bbox{r})
&=& C_{0}^{\rho}                            \rho_{0}^{2}
+  C_{0}^{\Delta\rho}                      \rho_{0}
\Delta\rho_{0}
+  C_{0}^{\tau}                            \rho_{0}
\tau_{0}
+  C_{0}^{J0}                               {J}_{0}^{2}
+  C_{0}^{J1}                          \bbox{J}_{0}^{2}
+  C_{0}^{J2}             \underline{\mathsf J}_{0}^{2}
+  C_{0}^{\nabla {J}}                      \rho_{0}
\bbox{\nabla}\cdot\bbox{J}_{0}
\nonumber \\
&+& C_{0}^{s}                           \bbox{s}_{0}^{2}
+  C_{0}^{\Delta{s}}                   \bbox{s}_{0}
\cdot\Delta\bbox{s}_{0}
+  C_{0}^{T}                           \bbox{s}_{0}
\cdot\bbox{T}_{0}
+  C_{0}^{j}                           \bbox{j}_{0}^{2}
+  C_{0}^{\nabla{j}}                   \bbox{s}_{0}
\cdot(\bbox{\nabla}\times\bbox{j}_{0}      )
+  C_{0}^{\nabla{s}}(\bbox{\nabla}\cdot\bbox{s}_{0}      )^2
+  C_{0}^{F}                           \bbox{s}_{0}
\cdot\bbox{F}_{0}       ,
\\ \label{chi1ph}
{\chi}_{1}(\bbox{r})
&=& C_{1}^{\rho}                           \vec{         \rho}^{\,2}
+  C_{1}^{\Delta\rho}                     \vec{         \rho}
\circ       \Delta\vec{         \rho}
+  C_{1}^{\tau}                           \vec{         \rho}
\circ             \vec{         \tau}
+  C_{1}^{J0}                             \vec{            J}^{\,2}
+  C_{1}^{J1}                             \vec{     \bbox{J}}^{\,2}
+  C_{1}^{J2}\underline{                  \vec{ {\mathsf J}}}^{\,2}
+  C_{1}^{\nabla {J}}                     \vec{         \rho}
\circ\bbox{\nabla}\cdot\vec{     \bbox{J}}
\nonumber \\
&+& C_{1}^{s}                              \vec{     \bbox{s}}^{\,2}
+  C_{1}^{\Delta{s}}                      \vec{     \bbox{s}}
\cdot \circ       \Delta\vec{     \bbox{s}}
+  C_{1}^{T}                              \vec{     \bbox{s}}
\cdot \circ~            \vec{     \bbox{T}}
+  C_{1}^{j}                              \vec{     \bbox{j}}^{\,2}
+  C_{1}^{\nabla{j}}                      \vec{     \bbox{s}}
\cdot \circ~(\bbox{\nabla}\times\vec{     \bbox{j}}        )
+  C_{1}^{\nabla{s}}
(\bbox{\nabla}\cdot                   \vec{     \bbox{s}}        )^2
+  C_{1}^{F}                              \vec{     \bbox{s}}
\cdot \circ~            \vec{     \bbox{F}}        ,
\enll
\end{widetext}
where $\times$ stands for the vector product of vectors in space,
$\circ$ stands for the scalar product of isovectors in isospace, and
other definitions closely follow those introduced in
Ref.~\cite{(Per04)}.

Quasilocal densities $\rho_k$, $\tau_k$, $\bbox{s}_k$, $\bbox{T}_k$,
$\bbox{j}_k$, $\bbox{F}_k$, $J_k$, $\bbox{J}_k$, and $\underline{\sf
J}_k$, are defined through the particle and spin non-local densities,
\begin{align}
  \rho_k(\bbox{r},\bbox{r}') =& \sum_{stt'} \hat{\rho}(\bbox{r}st,\bbox{r}'st') \hat{{\tau}}^k_{t't}, \\
  \bbox{s}_k(\bbox{r},\bbox{r}') =& \sum_{ss'tt'} \hat{\rho}(\bbox{r}st,\bbox{r}'s't') \hat{\bbox{\sigma}}_{s's}\hat{{\tau}}^k_{t't},
\end{align}
where $k$ runs from 0 to 3, $\hat{\bbox{\sigma}}$ and $\hat{\tau}^k
(k=1,2,3)$ are the Pauli matrices for spin and isospin,
respectively, and $\hat{\tau}^0_{t't} = \delta_{t't}$. The explicit
definitions and expressions in the cylindrical basis for the local
densities appearing in Eqs.~(\ref{chiph}) are given in
Appendix~\ref{sec:symmetry}. By varying the pnEDF with
respect to the density matrices, one obtains the {\particle}
mean-field Hamiltonian:
\begin{align}
\hat{h}(\bbox{r}'s't',\bbox{r}st)
&= \frac{\delta \overline{H}[\hat{\rho}]}
{\delta \hat{\rho}(\bbox{r}st,\bbox{r}'s't')} \nonumber \\
 & =  - \frac{\hbar^{2}}{2m}\delta(\bbox{r}-\bbox{r}')
\bbox{\nabla}\cdot  \bbox{\nabla}\,\delta_{s's}\delta_{t't}
\nonumber \\
& +   \hat{\Gamma}              (\bbox{r}'s't',\bbox{r}st)  +   \hat{\Gamma}_{\text{r}}(\bbox{r}'s't',\bbox{r}st),
\end{align}
where $\hat{\Gamma}$ is the HF potential and $\hat{\Gamma}_{\text{r}}$ is the rearrangement potential.

For the pnEDF depending on quasilocal densities only, such as in
Eq.~(\ref{chiph}), the HF Hamiltonian is a local differential
operator,
\be
\hat{{h}}(\bbox{r}'s't',\bbox{r}st) =
\delta (\bbox{r}-\bbox{r}')\hat{{h}}(\bbox{r};s't',st),
\ee
which has a simple isospin structure:
\be
\hat{h}(\bbox{r};s't',st) =
{h}_0(\bbox{r};s',s)\delta_{t't}+
\vec{h}(\bbox{r};s',s)\circ \hat{\vec{\tau}}_{t't}.  \label{eq:mfHamiltonian}
\ee
The isoscalar and isovector parts of the HF Skyrme Hamiltonian
can be written  in the compact form as,
\begin{align}\label{hatild1}
h_k(\bbox{r};s',s)
&=- \frac{\hbar^{2}}{2m}\bbox{\nabla}^2\delta_{s's}\delta_{k0}
+ U_k \delta_{s's}
+ \bbox{\Sigma}_k \cdot\hat{\bbox{\sigma}}_{s's} \nonumber \\
& + \frac{1}{2i}\big[\bbox{I}_k \delta_{s's}
+ ({\mathsf B}_k \cdot\hat{\bbox{\sigma}}_{s's})\big]\cdot\bbox{\nabla} \nonumber \\
& + \frac{1}{2i}\bbox{\nabla}\cdot\big[\bbox{I}_k \delta_{s's}
+ ({\mathsf B}_k \cdot\hat{\bbox{\sigma}}_{s's})\big]
\nonumber \\
& - \bbox{\nabla}\cdot\big[M_k \delta_{s's}
+ \bbox{C}_k \cdot\hat{\bbox{\sigma}}_{s's}\big]\bbox{\nabla} \nonumber \\
& - \bbox{\nabla}\cdot \bbox{D}_k \,\hat{\bbox{\sigma}}_{s's}\cdot \bbox{\nabla},
\end{align}
where $k=0,1,2,3$, and
\begin{align}\label{tenact}
({\mathsf B}\cdot\hat{\bbox{\sigma}})_a
=\sum_b{\mathsf B}_{ab}\hat{\bbox{\sigma}}^b ,
\end{align}
for $a=x,y,z$, denotes the $a$th component of a space vector.
By introducing the unit space tensor  ${\mathsfI}$ and the antisymmetric space
tensor $(\bbox{\epsilon}\cdot\bbox{J})_{ab} =
\sum_{c}\bbox{\epsilon}_{acb}\bbox{J}_c$,
the local real potentials can be written as:
\begin{subequations}\label{potall}
\begin{align}
U_k(\bbox{r})&= 2C^{\rho}_t\rho_k
+2C^{\Delta \rho}_t\Delta \rho_k  \nonumber \\
& + C^{\tau}_t\tau_k
+ C^{\nabla J}_t\bbox{\nabla}\cdot
\bbox{J}_k, \\
\bbox{\Sigma}_k(\bbox{r})&=2C^s_t\bbox{s}_k
+ 2(C^{\Delta s}_t  -C^{\nabla s}_t)\Delta
\bbox{s}_k  \nonumber \\
& - 2C^{\nabla s}_t\bbox{\nabla}\times
(\bbox{\nabla}\times\bbox{s}_k  ) \nonumber \\
& +C^T_t\bbox{T}_k
+ C^F_t\bbox{F}_k
+ C^{\nabla j}_t\bbox{\nabla}\times
\bbox{j}_k  , \\
\bbox{I}_k(\bbox{r})&= 2C^j_t\bbox{j}_k
+ C^{\nabla j}_t\bbox{\nabla}
\times\bbox{s}_k  ,\\
{\mathsf B}_k(\bbox{r})&= 2C^{J0}_tJ_k  {\mathsfI}
- 2C^{J1}_t\bbox{\epsilon}\cdot\bbox{J}_k  \nonumber \\
& + 2C^{J2}_t\underline{\mathsf J}_k
+ C^{\nabla J}_t\bbox{\epsilon}\cdot\bbox{\nabla}\rho_k   ,\\
M_k(\bbox{r})&=C^{\tau}_t\rho_k  , \\
\bbox{C}_k(\bbox{r})&=C^T_t\bbox{s}_k  ,   \\
\bbox{D}_k(\bbox{r})&=C^F_t\bbox{s}_k.
\end{align}
\end{subequations}
All coupling constants $C_t$ in Eqs.\ (\ref{potall})
are taken with $t=0$ for $k=0$ (isoscalars), and with $t=1$ for $k=1,
2, 3$ (isovectors).\footnote{Note that traditionally (cf.~Ref.~\cite{(Per04)})
the same symbol $t$ denotes either the s.p.\ isospin coordinate
$t=\pm\thalf$ or isoscalar/isovector coupling constants ($t=0/t=1$).
The meaning of $t$ thus has to be inferred from the context in which
it is used.}

The resulting HF equation can be written as a self-consistent  eigenvalue problem,
\begin{equation}
\int d^3 \bbox{r} \sum_{st}
\hat{h}(\bbox{r}'s't',\bbox{r}st) V_k^\ast(\bbox{r}st) = \varepsilon_k V^\ast_k(\bbox{r}'s't'),
\label{eig}
\end{equation}
which is solved by filling the lowest $A$ s.p.\ orbits in the
density matrix (\ref{eq:rho}).

\section{MODEL STUDY \label{sec:model}}
Recently,  two computer
codes capable of  solving the self-consistent equations for the isospin-invariant
pnEDFs with the p-n mixing,  have been developed in parallel.  The recent study~\cite{(Sat13c)} describes the scheme
based on the
code {\sc hfodd}~\cite{(Sch12),*(Sch14)}, which can treat symmetry-unrestricted nuclear shapes.
In this work, we present the implementation  based on the code {\sc hfbtho}, which assumes
axial and time-reversal symmetries. The two codes complement each other
in that {\sc hfodd} is more general whereas {\sc hfbtho} is much faster,
and thus they have different scopes and application ranges.
While {\sc hfbtho} can employ the transformed
oscillator basis that is particularly useful for weakly bound nuclei,
the focus of the present work is to benchmark {\sc hfbtho}  with 
{\sc hfodd}; hence, we shall use the standard
harmonic oscillator basis.

As in Ref.~\cite{(Sat13c)}, we diagonalize the s.p.\ Routhian,
\begin{equation}\label{eq:Routhian}
\hat{h}' = \hat{h} - \vec{\lambda}\circ\hat{\vec{t}},
\end{equation}
where $\hat{h}$ is given by Eq.~(\ref{eq:mfHamiltonian}) and contains
kinetic, Skyrme-pnEDF, and Coulomb-energy terms. The isocranking
term~\cite{(Sat01a),*(Sat01b)},
$-\vec{\lambda}\circ\hat{\vec{t}}$, depends on the isocranking frequency
(isovector Fermi energy) $\vec{\lambda}$ and the s.p.\
isospin operator  $\hat{\vec{t}}=\hat{\vec{\tau}}/2$.

For systems obeying the time-reversal symmetry, $\bra\hat{t}_y\ket$
vanishes~\cite{(Per04)} and the rotation in isospace  is described by
a two-dimensional isocranking, that is,
\begin{equation}\label{crankh}
-\vec{\lambda}\circ\hat{\vec{t}} =  - \lambda_x \hat{t}_x - \lambda_z \hat{t}_z.
\end{equation}
The isocranking frequencies, $\lambda_x$ and $\lambda_z$, can be varied to control the
isospin of the system. Following the methodology developed in Ref.~\cite{(Sat13c)},
they are parametrized as
\begin{align}\label{shifted}
  \lambda_z = \lambda'\cos\theta' + \lambda_{\text{off}}, \quad \lambda_x = \lambda' \sin\theta',
\end{align}
and the isocranking tilting angle $\theta'$ is varied between $0^\circ$ and $180^\circ$,
that is, the isovector Fermi energy on the $\lambda_x$--$\lambda_z$
plane moves along a shifted semicircle.
In this work, numerical calculations were performed for the $A=78$
IASs with $T\simeq11$. We
used the Skyrme EDF parametrization SkM*~\cite{(Bar82)} and
the s.p.\ basis space consisting of $N_{\rm sh} = 16$ spherical
harmonic-oscillator (HO) shells. Fore more details on the parameters employed, see Sec.~\ref{sec:comparison}.

In the absence of the Coulomb interaction, choosing $\lambda' = 21$\,MeV,  $\lambda_{\text{off}} = 0$\,MeV, and varying $\theta'$ from
$0^\circ$ to $180^\circ$, generates all the $A=78$ and $T=11$ IASs.
The angles $\theta'=0^\circ$, $90^\circ$, and $180^\circ$ correspond to the
HF solutions for $^{78}$Ni ($T_z=11$),
$^{78}$Y ($T_z=0$ in the odd-odd system), and  $^{78}$Sn ($T_z=-11$),
respectively. Our example involves very exotic nuclei, including those
beyond the proton dripline. We find this case interesting because the nuclei at both ends of the isobaric chain are heavy and  doubly magic, thus spherical.

As discussed in Ref.~\cite{(Sat13c)}, with the Coulomb term off, the value of
 $\lambda'$ is roughly equal to the absolute value of
difference between the proton and neutron Fermi energies in $^{78}$Ni
or $^{78}$Sn, $|\lambda_n - \lambda_p| = 21.18$\,MeV. Then, the
isocranking term makes the Fermi energies of neutrons and protons
almost equal. In the presence of the Coulomb interaction, however, a large asymmetry between $|\lambda_n - \lambda_p|$ develops between $^{78}$Ni
(12.31\,MeV) and $^{78}$Sn (33.62\,MeV).
Therefore, to
offset the difference of Fermi energies at $\theta'=0^\circ$ and
$180^\circ$ with Coulomb interaction present,
we  set the values:
\begin{subequations}\label{lambdas}
\begin{align}
\lambda'=\frac{1}{2}\{|\lambda_n-\lambda_p|(^{78}{\rm Ni})+|\lambda_n-\lambda_p|(^{78}{\rm Sn})\}, \\
\lambda_{\rm off}=\frac{1}{2}\{|\lambda_n-\lambda_p|(^{78}{\rm Ni})-|\lambda_n-\lambda_p|(^{78}{\rm Sn})\}. 
\end{align}
\end{subequations}

Using these expressions, $\lambda' = 22.94$\,MeV and $\lambda_{\text{off}} = -10.92$\,MeV.

\begin{figure}[htbp]
\includegraphics[width=0.9\columnwidth]{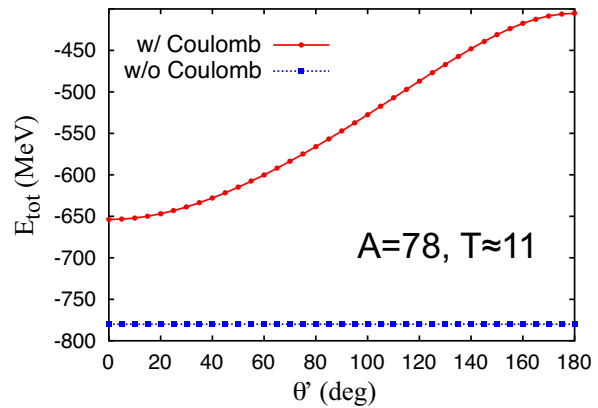}
\caption{(Color online)
Total HF energy of the $A=78$, $T\approx 11$ IASs as a function of  $\theta'$ with (solid line) and without (dashed line) the isospin-symmetry-breaking Coulomb term.
\label{fig:totalenergy}}
\end{figure}

Figure~\ref{fig:totalenergy} shows that in the absence of
the Coulomb interaction, the total energy is independent of
$\theta'$. This should be the case, as the pnEDF is isospin-invariant
and thus the energy must be independent of the direction of the isospin vector.
This also turned out to be an important test on the derived
expressions and numerical code, as different terms of
pnEDF become active for different values of $\theta'$. For
$\theta'=0^\circ$ and $180^{\circ}$, solutions are unmixed and the
densities are block-diagonal in neutron and proton subspaces. At
intermediate values of $\theta'$, the solutions are p-n mixed. For the
special case of $\theta'=90^{\circ}$, proton and neutron densities are equally mixed.
When the Coulomb interaction is turned on, the total energy increases
with $\theta'$ (Fig.~\ref{fig:totalenergy}), because more and more
protons replace neutrons and the Coulomb repulsion grows.

\begin{figure}[htbp]
\includegraphics[width=0.8\columnwidth]{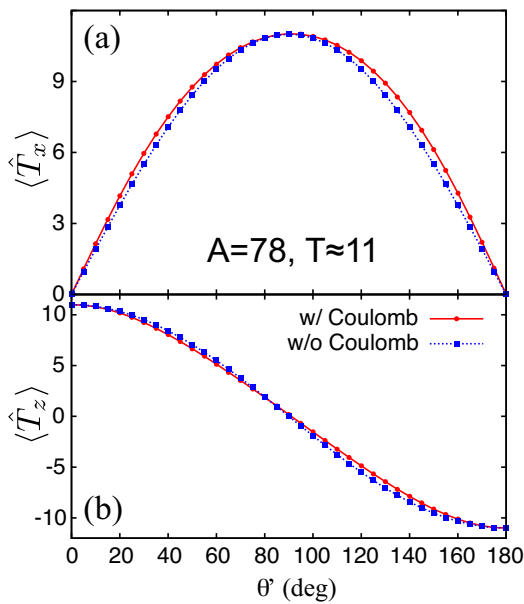}
\caption{(Color online)
Similar as in  Fig.~\protect\ref{fig:totalenergy}, but for the
expectation values of $\bra \hat{T}_x\ket$ (a) and $\bra
\hat{T}_z\ket$ (b). 
\label{fig:TxTz}}
\end{figure}

The degree of p-n mixing  can be
directly inferred from the expectation values of $\bra\hat{T}_x\ket$
plotted in Fig.~\ref{fig:TxTz}(a). As expected, the p-n mixing
increases with $\theta'$ and reaches its maximum value for
$\theta'=90^{\circ}$, and then drops again.  In
Fig.~\ref{fig:TxTz}(b), we show  $\bra\hat{T}_z\ket$ and it is seen
 that the values of $\theta'=0^\circ$, $\sim$$90^\circ$, and
$180^\circ$ do correspond to $^{78}$Ni,  $^{78}$Y, and  $^{78}$Sn,
respectively. The behavior of $\bra\hat{T}_z\ket$ and $\bra\hat{T}_x\ket$ weakly depends on whether the Coulomb term
is included or not. This is entirely due to our
choice of the shifted semicircle (\ref{shifted}), whereupon the
linear constraint $\lambda_{\text{off}} \hat{t}_z$ absorbs the major
part of the isovector component of the Coulomb interaction.

\begin{figure}[htbp]
\includegraphics[width=0.9\columnwidth]{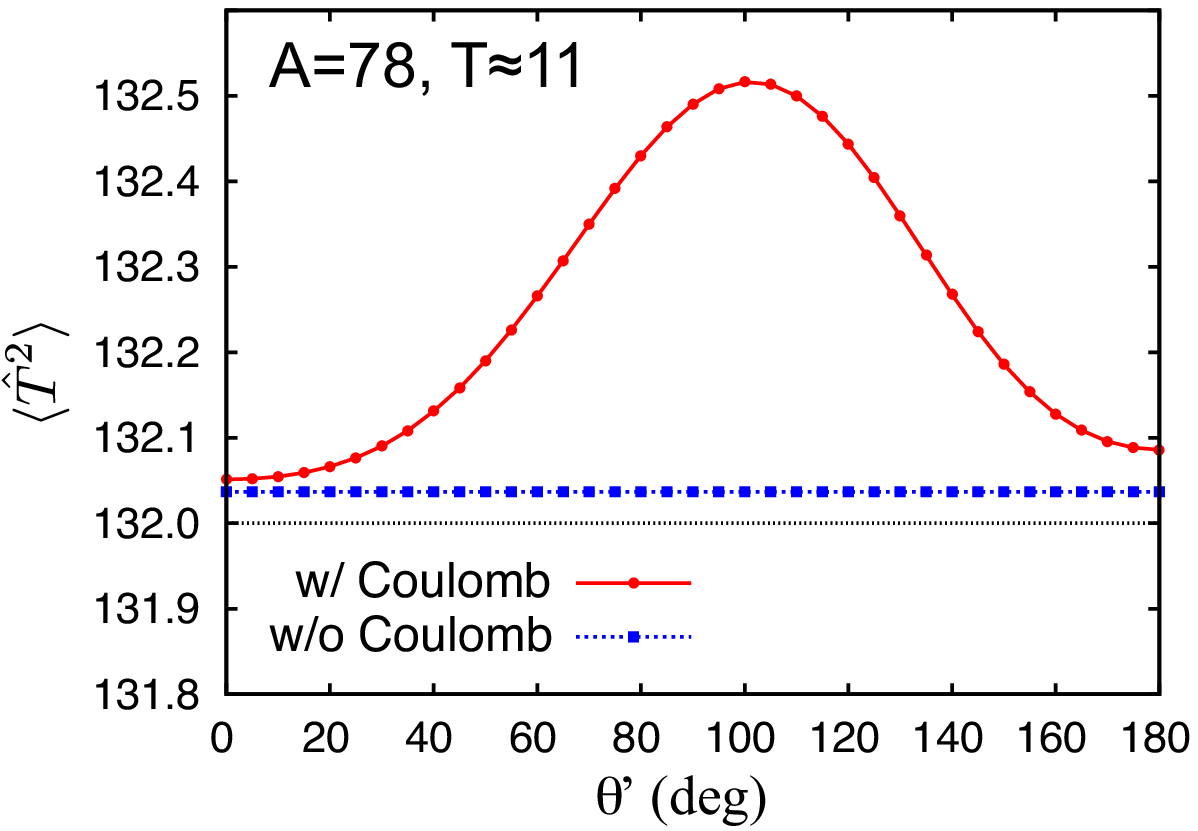}
\caption{(Color online)
Similar as in Fig.~\protect\ref{fig:totalenergy}, but for $\bra \hat{T}^2 \ket$.
\label{fig:T2}}
\end{figure}

The Coulomb interaction breaks  isospin  and thus induces
the isospin mixing in the HF wave function. To illustrate this,
Fig.~\ref{fig:T2} shows  the average value of $\bra\hat{T}^2\ket$
for the converged HF solutions. For
the considered case of the $T=11$ systems, $\bra\hat{T}^2\ket$ should be exactly equal to $11\times 12=132$ in the absence of
isospin mixing.
However, as shown in Fig.~\ref{fig:T2},  even in the absence
of the Coulomb interaction, $\bra\hat{T}^2\ket$ slightly deviates from this value. At the origin of this effect is the spurious isospin mixing~\cite{(Eng70),(Aue83),(Sat09)}. Indeed, within the mean-field
approximation, the isospin symmetry is broken spontaneously as the HF wave function is not an isospin eigenstate. However, since the Skyrme EDF is isospin covariant \cite{(Car08),(Roh10)}, the HF solutions corresponding to different orientations in the isospin space are
degenerate in energy. While the neutron-proton mixing changes with the angle $\theta$, 
$\bra\hat{T}^2\ket$ must remain the  same in the absence of the Coulomb interaction. 
In the presence of the
Coulomb term, the isospin mixing is very
small in the isospin-stretched $|T_z|=11$ configurations (for
$\theta'=0^{\circ}$ and $180^{\circ}$) and reaches its maximum around
$\theta'=90^{\circ}$ for  $T_z=0$  \cite{(Sat09),(Sat10)}.

\begin{figure}[htbp]
\includegraphics[width=0.9\columnwidth]{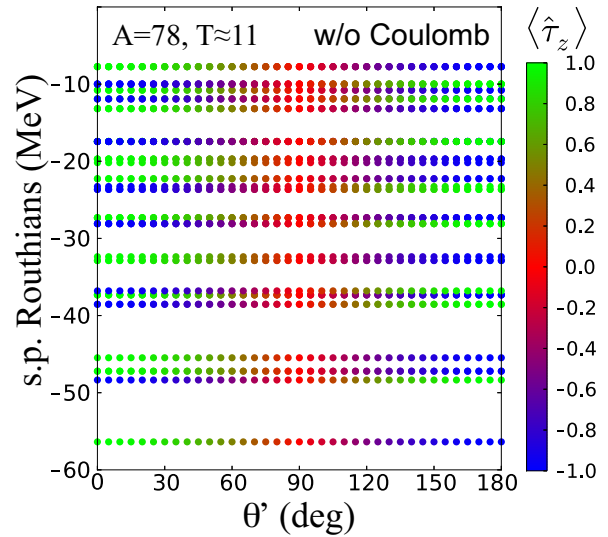}
\caption{(Color online)
Single-particle Routhians as functions of  $\theta'$ for
$T\approx 11$ configurations in the $A=78$ systems.
The Coulomb interaction is not included.
Points are colored according to the s.p.\ expectation values of $\bra \hat{\tau}_z\ket$.
At $\theta'=0^\circ$,
neutron and proton states are plotted up to 1$g_{7/2}$ and 2$p_{1/2}$, respectively.
\label{fig:spenergywocoul}}
\end{figure}
\begin{figure}[htbp]
\includegraphics[width=0.9\columnwidth]{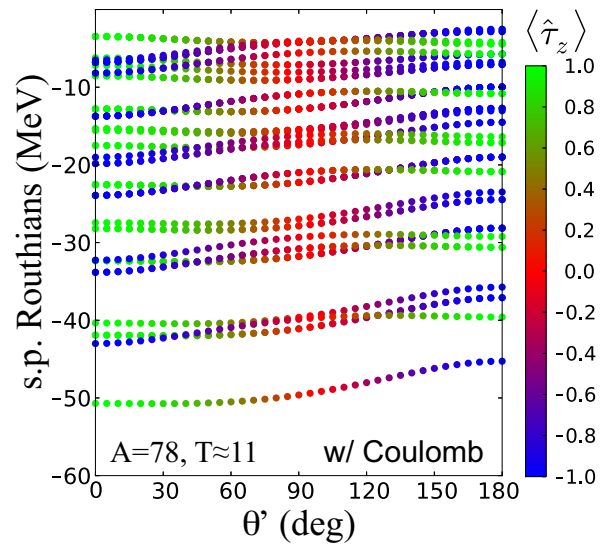}
\caption{(Color online)
Similar to Fig.~\ref{fig:spenergywocoul}, but with
the Coulomb interaction included.
\label{fig:spenergywcoul}}
\end{figure}

The s.p.\ Routhians as functions of $\theta'$ are shown in
Figs.~\ref{fig:spenergywocoul} (without Coulomb) and \ref{fig:spenergywcoul} (with Coulomb). Eleven
spherical neutron levels and seven proton levels are occupied at
$\theta'=0^\circ$, and the neutron and proton Fermi energies are
shifted in such a way that the gaps in the s.p.\ spectra appear at $A=78$
around $-15$\,MeV (Fig.~\ref{fig:spenergywocoul}) and $-10$\,MeV
(Fig.~\ref{fig:spenergywcoul}). Our  choice of $\vec{\lambda}$ guarantees that,
in the presence of the Coulomb interaction,
the s.p.\ Routhians near the Fermi surface do not cross as functions of $\theta'$; this  would have caused a
drastic structural changes of the mean-field and made
the adiabatic
tracing of the $T\simeq11$ IAS as a function of $\theta'$ extremely difficult.
At $\theta'=0^\circ$ and $180^{\circ}$, the s.p.\ states
have pure values of $\bra\hat{t}_z\ket=\pm\thalf$. At $\theta' \sim
90^{\circ}$, most of the s.p.\ Routhians  have $ \bra\hat{t}_z\ket$ close to
zero, that is, they are fully p-n mixed.

\begin{figure}[htbp]
\includegraphics[width=0.9\columnwidth]{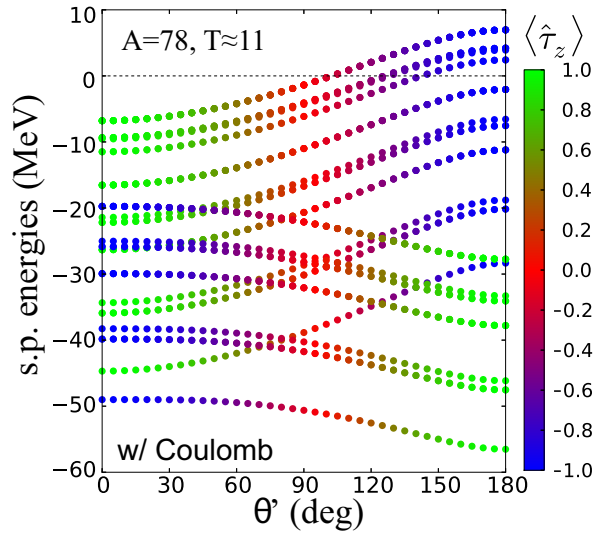}
\caption{(Color online)
Similar to Fig.~\ref{fig:spenergywocoul} but for the s.p.\ energies
with the Coulomb interaction included. Only the occupied states are
plotted, that is, at $\theta'=0^\circ$, neutron and proton states are
plotted up to 1$g_{9/2}$ and 1$f_{7/2}$ shells, respectively.
\label{fig:spenergywcoul-lxzremoved}}
\end{figure}

Figure~\ref{fig:spenergywcoul-lxzremoved} displays s.p.\ HF energies,
that is, s.p.\ Routhians with the isocranking term removed.
Note that these are not eigenvalues but the average values of the HF Hamiltonian,
calculated for states that are eigenstates of the Routhian (\ref{eq:Routhian}).
With increasing $\theta'$, owing to the increasing Coulomb field,
s.p.\ states that increase proton (neutron) component gradually
increase (decrease) in energy.

\begin{figure}[htbp]
\includegraphics[width=0.8\columnwidth]{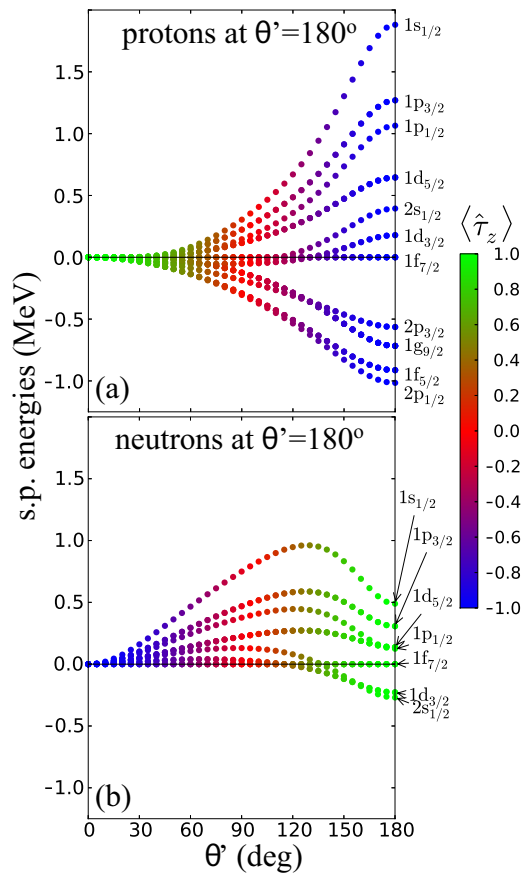}
\caption{(Color online)
Evolution of s.p.\ energies of  occupied HF states (with Coulomb) originating from
proton (top) and neutron (bottom) shells at $\theta'=180$ with respect
to the energy of the 1$f_{7/2}$ shell, and relative to the corresponding  values at
$\theta'=0^\circ$.  That is, the energies plotted are
$[\varepsilon_i(\theta') - \varepsilon_{f_{7/2}}(\theta')] -
[\varepsilon_i(0^\circ) - \varepsilon_{f_{7/2}}(0^\circ)]$.
\label{fig:spenergywcoul-lxzremoved-f7}}
\end{figure}

To better visualize the relative shifts of s.p.\ levels with  $\theta'$, in Fig.~\ref{fig:spenergywcoul-lxzremoved-f7}  we show s.p.\
energies relative to the energy of the  1$f_{7/2}$ shell.
The figure nicely illustrates  the effect of the Coulomb interaction
on the proton components of the s.p.\ orbits: the
relative level shifts correlate  with their binding energies and $\ell$ values~\cite{(Shl78),(Naz96)}. Indeed, the
deeply (loosely) bound levels, which have smaller (larger) rms radii
and thus experience stronger (weaker) Coulomb repulsion, are shifted
up (down) in energy relative to the high-$\ell$ 1$f_{7/2}$ shell.

Some of the calculated $A=78$, $T\simeq11$ IASs are predicted to appear beyond the proton
drip line. As seen in Fig.~\ref{fig:spenergywcoul-lxzremoved}, energy
of the 1g$_{9/2}$ level (which is neutron at $\theta'=0^\circ$ and
proton at $\theta'=180^\circ$) becomes positive at around
$\theta'=100^\circ$, where $\langle\hat{T}_z\rangle \approx -1.5$.
At $\theta'=180^\circ$, energies of the 1$g_{9/2}$, 1$f_{5/2}$,
2$p_{1/2}$, and 2$p_{3/2}$ shells are  positive. However, all these states are well localized by the Coulomb barrier, and thus
correspond to narrow resonances, whose energies can be reasonably well
described within the HO basis expansion~\cite{(Naz96)}.

\begin{figure}[htbp]
\includegraphics[width=0.8\columnwidth]{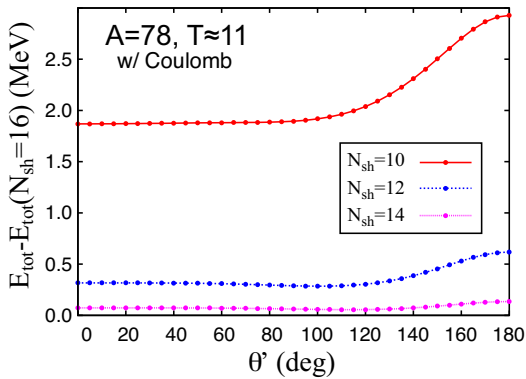}
\caption{(Color online)
The total HF energy (with Coulomb) for $A=78$, $T\approx 11$  calculated
 with  $N_{\rm sh}=10,
12, $ and $14$ HO shells relative to that  with
$N_{\rm sh}=16$ shells as a function of $\theta'$.
\label{fig:energy-Nsh}}
\end{figure}
\begin{figure}[htbp]
\includegraphics[width=1.0\columnwidth]{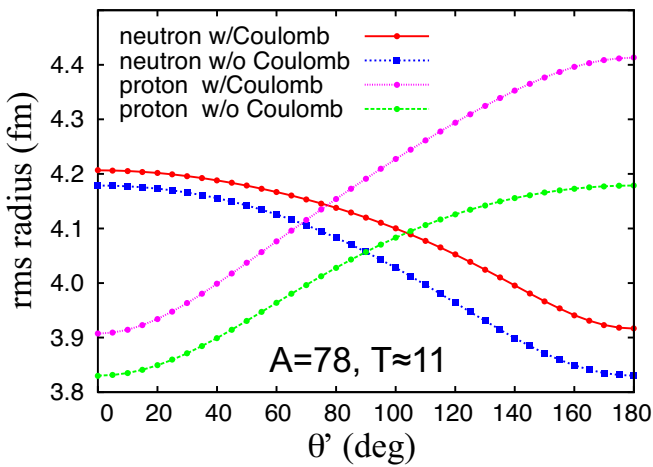}
\caption{(Color online)
Neutron and proton rms radii as functions of $\theta'$.
\label{fig:radius}}
\end{figure}
In Fig.~\ref{fig:energy-Nsh}, we show the convergence of the total
HF energy with respect to the total number of HO shells $N_{\rm sh}$.
Calculations using
$N_{\rm sh}=14$ are not yet completely converged, with the energy
difference between  $N_{\rm sh}=16$ and $N_{\rm sh}=14$ varying
around  73\,keV at $\theta'=0^\circ$ and
135\,keV at $\theta'=180^\circ$. Although it is expected
that by increasing $N_{\rm sh}$ one may still lower  the  energy, the
change is expected to be less than 100\,keV, and the results
presented in this study are not expected to change significantly. A
delayed convergence beyond $\theta'=140^\circ$ shows that the higher HO
shells are more important at $\theta'=180^\circ$ than at
$\theta'=0^\circ$; that is, a larger model space is required for the
description of the unbound proton resonances \cite{(Kru04)}. Nevertheless,
as seen in Fig.~\ref{fig:radius}, even near $\theta'=180^\circ$ no
sudden increase of proton rms radii is obtained.

\begin{figure*}[htbp]
\includegraphics[width=0.7\textwidth]{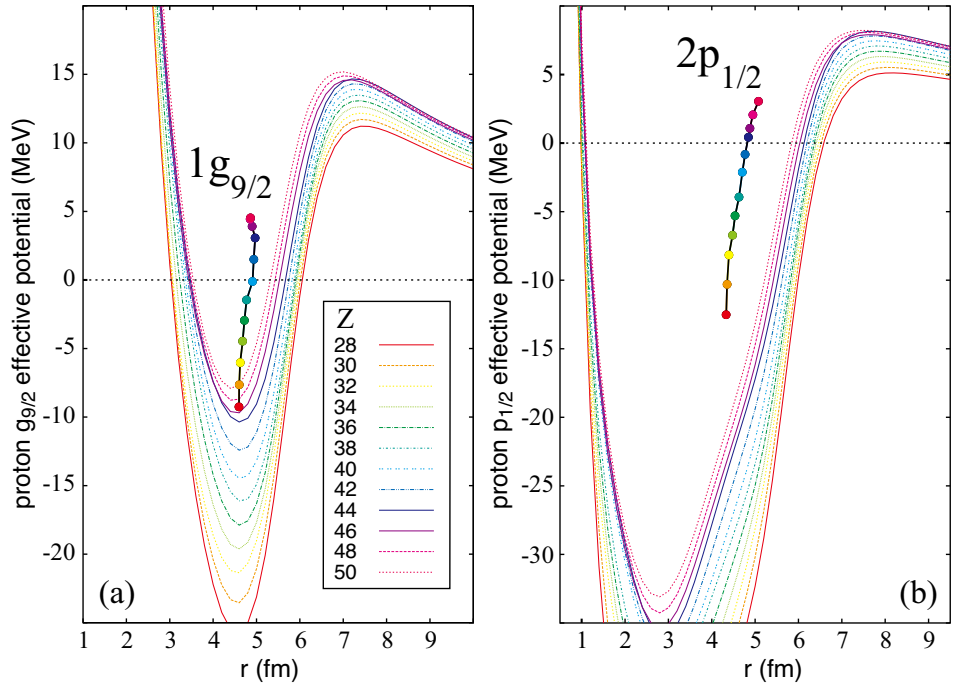}
\caption{(Color online)
Lines: proton effective HF potential (with Coulomb and
 centrifugal terms included), calculated for the $A=78$,
$T\simeq|T_z|$
isobars for $\ell=4$ (a) and $\ell=1$ (b). Dots: rms radii and s.p.\ energies of the proton
1$g_{9/2}$ (a) and 1$p_{1/2}$  states.
\label{fig:protong9potential}}
\end{figure*}

To investigate properties of the unbound proton orbits, for the ground states of
$A=78$, $T\simeq|T_z|$ nuclei, we performed the {\sc hfbrad}
\cite{(Ben05b)} calculations (without p-n mixing). In
Fig.~\ref{fig:protong9potential}, we
show results obtained for the 1$g_{9/2}$ and 2$p_{1/2}$ proton
states. For each $A=78$ isobar, a dot is placed at the
values of s.p.\ energies and radii, and lines show standard total
effective HF proton potentials.
The total
effective HF proton potential consists of the standard central, spin-orbit, centrifugal, 
and Coulomb terms.
The proton 1$g_{9/2}$
orbit in $^{78}$Zr is bound, and in $^{78}$Mo it becomes slightly unbound. This
result is consistent with the experimental observation, whereby the
last bound nucleus of the $A=78$ isobaric chain, which is
experimentally known, is $^{78}$Zr $(T_z=-1)$. The rms radii of the
proton 1$g_{9/2}$ orbits are about 5\,fm, and the s.p.\ wave
functions are still localized, even if the orbits become unbound.
This is because the  1$g_{9/2}$  and 2$p_{1/2}$ protons
occupy states well below the potential barrier, which pushes the proton continuum up in energy, thus effectively extending the range of nuclear landscape into the proton-unstable region \cite{(Ver00),(Ols13),*(Ols13a)}.

It is worth noting that the 2$p_{1/2}$ orbit, 
which has a small centrifugal barrier, is bound up to around ($\theta'=125^\circ, \langle\hat{T}_z\rangle\approx-5.7$) in Fig.~\ref{fig:spenergywcoul-lxzremoved}. This is consistent
with Fig.~\ref{fig:protong9potential} that shows that the s.p. energy of 2$p_{1/2}$ is unbound in $^{78}$Ru $(\langle\hat{T}_z\rangle=-5)$.
In the presence of p-n mixing,
the proton components of the s.p.\ states are smaller than those in
the pure proton states, and this effectively reduces the repulsive
Coulomb energies of the 1$g_{9/2}$ orbits.

\section{Benchmarking with {\sc\bf hfodd} \label{sec:comparison}}

To demonstrate that the isospin-invariant formalism has been
properly implemented, we provide a detailed comparison between  the {\sc hfodd}~\cite{(Sat13c)} and {\sc
hfbtho} frameworks. This benchmarking is meaningful as the two pnEDF codes were developed
independently and have fairly different structures. In particular, the HF equations in
{\sc hfodd}~\cite{(Sch12),*(Sch14)}  are solved in three-dimensional
Cartesian basis while {\sc hfbtho} employs the two-dimensional  cylindrical basis.

Calculations were performed for the $A=40$, $T\simeq8$ deformed IASs
with  SkM*   EDF parametrization using the   s.p.\ basis space
 of $N_{\rm sh} = 10$. The oscillator length was
assumed to be  $b$ = 1.697626\,fm.
The mass constant in Eq.~(\ref{enden})
was fixed at $\hbar^2/2m = 20.73$\,MeV\,fm$^2$. As far as integration is concerned,  we used $N_{\rm GH} = 26$ Gauss-Hermite nodes for each Cartesian coordinate in  {\sc hfodd},  whereas in
{\sc hfbtho}, the numbers of Gauss-Hermite ($\rho$-direction) and Gauss-Laguerre ($z$-direction) nodes were assumed to be equal: $N_{\rm GH} = N_{\rm GL}=40$. In addition, in {\sc
hfbtho}, the number of Gauss-Legendre nodes used in the integration
of the direct Coulomb field was set to 80, and the Coulomb length
scale  was taken to be $L=50$\,fm. This set of parameters was recommended as a
default value in the latest version of the {\sc hfbtho}, as it provides
a sufficient precision on the direct Coulomb energy~\cite{(Sto13)}.
Without  Coulomb, the isocranking frequency was set to
$\lambda'= 27.092394$\,MeV and $\lambda_{\text{off}} = 0$\,MeV. With
Coulomb, we took the values $\lambda'=28.613615$\,MeV
and $\lambda_{\text{off}} = -6.010741$\,MeV.

\begin{table}
\caption{\label{table:0deg}
Benchmarking of {\sc hfodd} with {\sc hfbtho} for  the deformed $A=40, T\simeq 8$ IASs using  SkM*  EDF ($N_{\rm sh} = 10$), and $\theta'=0^\circ$ ($^{40}$Mg). Other parameters are:
$\lambda' = 27.092394$\,MeV, $\lambda_{\rm off} = 0$\,MeV (without Coulomb),
and $\lambda' =28.613615$\,MeV, $\lambda_{\rm off} =  -6.010741$\,MeV (with Coulomb). Shown are various contributions to the total binding energy $E_{\rm tot}$ (in MeV), r.m.s. radii (in fm), expectation values of $T^2, T_z$, and $T_x$, and the quadrupole deformation $\beta_2$.
The digits which do not coincide in {\sc hfodd} and {\sc hfbtho} are marked in bold.
}
\begin{ruledtabular}
\begin{tabular}{c|cccc}
 & {\sc hfodd} & {\sc hfbtho}  & {\sc hfodd} & {\sc hfbtho}  \\ \hline
                             & \multicolumn{2}{c}{Without Coulomb}
                                & \multicolumn{2}{c}{With Coulomb} \\
                              \\[-8pt]
$E_{\rm tot}$          & -303.425{\bf 19} & -303.425{\bf 20}  & -276.4764{\bf 1} & -276.4764{\bf 3} \\
$E_{\rm kin}^{({\rm n})}$  & 498.44846{\bf 6} & 498.44846{\bf 4} &  495.539{\bf 29} & 495.539{\bf 30} \\
$E_{\rm kin}^{({\rm p})}$  & 175.37176{\bf 4} & 175.37176{\bf 2} &  171.3020{\bf 5} & 171.3020{\bf 6} \\
$E_{\rm pot}$         &-977.2454{\bf 2} & -977.2454{\bf 3}   & -970.0992{\bf 3} & -970.0992{\bf 6}\\
$E_{\rm SO}$            & -34.35790{\bf 3} & -34.35790{\bf 5}  & -33.18481{\bf 2} & -33.18481{\bf 6} \\
$E_{\rm Cou}^{({\rm dir})}$  & & & 30.920{\bf 704} & 30.920{\bf 697}  \\
$E_{\rm Cou}^{({\rm exc})}$ & & & -4.139228 & -4.139228 \\
$r_{\rm rms}^{({\rm n})}$   & 3.697718 & 3.697718   & 3.709975 &  3.709975 \\
$r_{\rm rms}^{({\rm p})}$   & 3.176356 & 3.176356  & 3.217587 & 3.217587  \\
$T^2$                        & 72.022743 & 72.022743    & 72.023123 & 72.023123 \\
 $T_z$                        & 8.000000 & 8.000000   & 8.000000 &  8.000000  \\
$T_x$                        & 0.000000 & 0.000000    & 0.000000 & 0.000000  \\
$\beta_2$                      &  0.304201 & 0.304201   & 0.311518 &  0.311518
\end{tabular}
\end{ruledtabular}
\end{table}

\begin{table}
\caption{\label{table:90deg}
Similar to Table~\ref{table:0deg} but for  $\theta'=90^\circ$ ($^{40}$Ca).}
\begin{ruledtabular}
\begin{tabular}{c|cccc}
 & {\sc hfodd} & {\sc hfbtho}  & {\sc hfodd} & {\sc hfbtho}  \\ \hline
                             & \multicolumn{2}{c}{Without Coulomb}
                                & \multicolumn{2}{c}{With Coulomb} \\
                              \\[-8pt]
$E_{\rm tot}$           &  -303.425{\bf 19} &  -303.425{\bf 20}  & -234.4{\bf 29} & -234.4{\bf 19} \\
$E_{\rm kin}^{({\rm n})}$  & 336.91011{\bf 5} & 336.91011{\bf 3} & 333.{\bf 68} & 333.{\bf 71} \\
$E_{\rm kin}^{({\rm p})}$   & 336.91011{\bf 5} & 336.91011{\bf 3}  & 318.7{\bf 04} & 318.7{\bf 13} \\
$E_{\rm pot}$           & -977.2454{\bf 20} & -977.2454{\bf 3}   & -954.{\bf 79} & -954.{\bf 83} \\
$E_{\rm SO}$              & -34.35790{\bf 3} &  -34.35790{\bf 5}     & -31.{\bf 49} & -31.{\bf 51} \\
$E_{\rm Cou}^{({\rm dir})}$  & & & 75.10{\bf 5} & 75.10{\bf 6} \\
$E_{\rm Cou}^{({\rm exc})}$  & & & -7.1252{\bf 2} & -7.1252{\bf 5} \\
$r_{\rm rms}^{({\rm n})}$   & 3.549360 & 3.549360   & 3.59{\bf 30} & 3.59{\bf 28} \\
$r_{\rm rms}^{({\rm p})}$   & 3.549360 & 3.549360   & 3.6351{\bf 9} & 3.6351{\bf 2} \\
$T^2$                        & 72.022743 & 72.022743     & 72.1{\bf 49} &  72.1{\bf 55}\\
$T_z$                        & 0.000000  & 0.000000    & 0.156{\bf 49}  &  0.156{\bf 52}\\
$T_x$                        & 8.000000  & 8.000000     &  8.005{\bf 1} & 8.005{\bf 4} \\
$\beta_2$                      &  0.304201   & 0.304201   & 0.318{\bf 4}  & 0.318{\bf 2}
\end{tabular}
\end{ruledtabular}
\end{table}

The benchmarking results for the   deformed case are  shown in Tables \ref{table:0deg} and  \ref{table:90deg} for $\theta'=0^\circ$ and $90^\circ$, respectively. In the absence of the Coulomb term, the difference in the total energy $E_{\rm tot}$ is
less than  20\,eV, and the total isospin $\bra\hat{T}^2\ket$
agrees up to the
sixth decimal place. With the inclusion of the Coulomb term, the agreement is slightly reduced but is still excellent. A comparison between {\sc hfbtho}
and {\sc hfodd}
was also performed for the spherical $A=48$, $T\simeq4$ IASs and the results are presented
in Table  \ref{table:90degsph} for $\theta'=90^\circ$ where the differences between the
two codes are  largest. In this case, it is found that
deviation in the total energy is about 20\,eV.

It is to be noted that both {\sc hfodd} and {\sc hfbtho}  use the same number of basis harmonic oscillator states. Moreover, 
as it has been 
demonstrated previously \cite{(Sto13)}, using a sufficient number of quadrature points in {\sc hfbtho} and {\sc hfodd}, the results of both solvers agree  with a high accuracy of  several eV.
The differences between the two codes 
with the Coulomb potential turned on, can be traced back to different techniques used to compute the direct Coulomb field: the solver  {\sc hfodd}, 
uses a more accurate Green's function approach.  The  benchmark examples discussed in this section  demonstrate that the p-n mixing has been implemented correctly in both  codes. 
Presently, we are in process of implementing the
cylindrical Green's function treatment of the Coulomb potential  into {\sc hfbtho}  and it is expected that the agreement between the two codes will further improve.

\begin{table}
\caption{\label{table:90degsph}
Similar to in Table~\ref{table:0deg} but  for the spherical  $A=48, T\simeq 4$ IASs and $\theta'=90^\circ$ ($^{48}$Cr). Other parameters are:
$\lambda' = 11.0$\,MeV, $\lambda_{\rm off} = 0$\,MeV (without Coulomb),
and $\lambda' =12.0$\,MeV, $\lambda_{\rm off} =  -8.0$\,MeV (with Coulomb).
}
\begin{ruledtabular}
\begin{tabular}{c|cccc}
 & {\sc hfodd} & {\sc hfbtho}  & {\sc hfodd} & {\sc hfbtho}  \\ \hline
                             & \multicolumn{2}{c}{Without Coulomb}
                                & \multicolumn{2}{c}{With Coulomb} \\
                              \\[-8pt]
 $E_{\rm tot}$          & -491.2437{\bf 06} & -491.2437{\bf 24}   &  -389.84{\bf 54}  & -389.84{\bf 38} \\
$E_{\rm kin}^{({\rm n})}$   & 422.9315{\bf 2} &  422.9315{\bf 8}  &   415.{\bf 69} & 415.{\bf 71}  \\
$E_{\rm kin}^{({\rm p})}$   & 422.931{\bf 52} &  422.931{\bf 43}  &   404.67{\bf 3} &  404.67{\bf 9} \\
$E_{\rm pot}$           & -1337.1067{\bf 5} & -1337.1067{\bf 3}    &  -1310.4{\bf 4} & -1310.4{\bf 6}\\\
$E_{\rm SO}$              & -36.73641{\bf 8} & -36.73641{\bf 7}  & -34.12{\bf 16} &  -34.12{\bf 40}\\
$E_{\rm Cou}^{({\rm dir})}$ & & &  109.37{\bf 74} & 109.37{\bf 81}\\
$E_{\rm Cou}^{({\rm exc})}$ & & &  -9.145{\bf 76} &  -9.145{\bf 81}\\
$r_{\rm rms}^{({\rm n})}$   & 3.4979{\bf 39} & 3.4979{\bf 40}  &  3.526{\bf 33} &  3.526{\bf 28}\\
$r_{\rm rms}^{({\rm p})}$   & 3.497939  & 3.497939   &  3.5778{\bf 7} &  3.5778{\bf 4}\\
$T^2$                        & 20.037818 & 20.037818   & 20.07{\bf 56}  & 20.07{\bf 71}\\
$T_z$                         & 0.00000{\bf 0} & 0.00000{\bf 3}     &  -0.0126{\bf 76} & -0.0126{\bf 63}\\
$T_x$                        &  4.000000 & 4.000000   &  4.002{\bf 64}  & 4.002{\bf 78}
\end{tabular}
\end{ruledtabular}
\end{table}

\section{SUMMARY AND OUTLOOK}\label{sec:summary}

The description of weakly bound  complex nuclei is a demanding task as it requires the understanding and control of three crucial aspects of the nuclear many-body problem: interaction, correlations, and coupling to the low-lying particle continuum. Here, the theoretical tool of choice is  nuclear density functional theory  based on the self-consistent EDF approach. The quest for a truly universal nuclear EDF is one of the main themes of theoretical nuclear structure research today.

The isospin channel of the nuclear EDF still remains largely unexplored. In the existing functionals, only isoscalar and $t_z$ components of the isovector densities are used as building blocks. In a completely isospin-invariant formalism, all three isovector density components should be considered: those correspond to  p-n mixed densities.
For heavy nuclei, possessing significant neutron excess, the omission of p-n mixed densities
can be justified as neutron and protons occupy different
shells. However, for lighter systems,
neutrons and protons usually occupy the same shell-model orbits and
p-n mixed densities are likely to appear.
Some limited experimental evidence suggests that the p-n fields
play a role  near the $N=Z$ line.

In the present work, we have developed a new Skyrme-EDF  approach with the inclusion of p-n mixed
densities \cite{(Per04)}. The expressions for the densities and HF  fields have been
worked out in the axial limit.
The present 2D {\sc hfbtho} implementation that includes  mixed p-n
densities and fields is fairly fast, and this  allows for systematic large-scale surveys.  The new
framework has  been  tested  in the HF
limit and it was benchmarked with  3D {\sc hfodd} spherical and deformed calculations  \cite{(Sat13c)}. The basic features of
the p-n mixed HF formalism have been investigated by studying the $A=78, T\approx 11$ IAS chain. In particular, we investigated the isospin breaking effects
and  stability of  solutions obtained for the proton-unbound systems.

The present work has been primarily devoted to the detailed test of  the newly developed
isospin-invariant density functional formalism. In the near future, we
intend  to perform realistic HFB calculations by including the
generalized pairing interaction in both isoscalar and isovector
channels in order to study the importance of the $T=0$   pairing densities
and fields on the structure of
nuclei close to the $N=Z$ line and the impact of p-n mixing on $\beta$ decays.

\begin{acknowledgments}
We would like to express our deep gratitude to our friend and
colleague, Mario Stoitsov, for his contributions in the initial stages
of this work. JAS would  like to acknowledge I.\ Maqbool
and P.A.\ Ganai for discussions. This work was supported by the U.S.\
Department of Energy under Contracts No.\ DE-FG02-96ER40963
(University of Tennessee), No.\ DE-SC0008499    (NUCLEI SciDAC
Collaboration), and No.\ DE-FG02-06ER41407 (JUSTIPEN, Japan-U.S.\
Theory Institute for Physics with Exotic Nuclei);  by the Academy of
Finland and University of Jyv\"askyl\"a within the FIDIPRO programme;
by the Polish National Science Center under Contract No.\
2012/07/B/ST2/03907; and by JSPS KAKENHI (Grants No. 20105003,
No. 24105006, No. 25287065, and No. 25287066). Computational resources were provided
through an INCITE award ``Computational Nuclear Structure" by the
National Center for Computational Sciences (NCCS) and the National
Institute for Computational Sciences (NICS) at Oak Ridge National
Laboratory.
A part of the numerical calculations were also carried out on SR16000
at the Yukawa Institute for Theoretical
Physics in Kyoto University and on the RIKEN Integrated Cluster of Clusters
(RICC) facility.

\end{acknowledgments}

\appendix

\section { CYLINDRICAL SYMMETRY \label{sec:symmetry}}
In the case of cylindrical symmetry, the third component $J_z$ of the
total angular-momentum is conserved and provides a good quantum
number $\Omega_k$. The HF s.p.\ wave functions in the axial limit
can be written as \cite{(Sto05)}
\begin{align}
V_k(\bbox{r}st)
=&
V_k^+ (rzt)
e^{i \Lambda^- \phi} \chi_{+1/2}(s) \nonumber \\
+&
V_k^- (rzt)
e^{i \Lambda^+ \phi} \chi_{-1/2}(s), \label{hfb-cyl-exp}
\end{align}
where $\Lambda^{\pm} = \Omega_k \pm 1/2$ and $(r,\phi,z)$ are the
cylindrical coordinates defining the three-dimensional position
vector, $\bbox{r} =( r\,\cos\phi, r\,\sin\phi, z )$ and $z$ is
the chosen symmetry axis. In the following section, the
expressions in the cylindrical coordinate basis under the axial
symmetry are provided for the local particle-hole
densities. In the second section, expressions of the
derivatives of the local densities, which are used in the evaluation
of the HF potentials, are given.

\subsection{Particle-hole densities}
We give the expression for particle density $\rho$,  kinetic density
$\tau$, spin density $\bbox{s}$, spin-kinetic density $\bbox{T}$,
current density $\bbox{j}$, and spin-current density ${\sf J}$.
Tensor-kinetic density $\bbox{F}$ is not used and its expression is
omitted.

\begin{itemize}
\item[(a)]  Scalar particle density
\begin{align}
\rho_{m}(\bbox{r}) = \rho_{m}(\bbox{r},\bbox{r}')\biggl|_{\bbox{r} = \bbox{r}'}
\end{align}
where $m$ takes values 0, 1, 2 and 3. Suffix 0 represents
isoscalar component of the density
and 1 to 3 are the isovector components.
Expressing the density in terms of HFB wavefunctions~(\ref{hfb-cyl-exp}), we obtain
\begin{align}
\rho_{m}(\bbox{r})= \sum_{tt'} \rho^{tt'}(rz)  \tau_{t't}^{m}
\end{align}
where
\begin{align}
\rho^{tt'}(rz) = \sum_{k}\biggl[ V^{+\ast}_{kt} V^{+}_{kt'}+V^{-\ast}_{kt} V^{-}_{kt'}\biggr],
\end{align}
and we use the abbreviation of HFB wave functions
\begin{align}
  V^{\pm}_{kt} \equiv V^{\pm}_k(rzt).
\end{align}
Isospin components of the particle density are given by
\begin{align}
\rho_{0}(\bbox{r}) =&\rho^{nn}(rz) +\rho^{pp}(rz) \nonumber  \\
                   =& \frac {1} {2} \biggr(\rho^{nn}(rz) +\rho^{pp}(rz)
+ c.c. \biggr), \label{eq:rho0} \\
\rho_{1}(\bbox{r}) =&\rho^{np}(rz) +\rho^{pn}(rz) \nonumber \\  =&  \rho^{np}(rz) + c.c., \\
\rho_{2}(\bbox{r}) =&i[\rho^{np}(rz) -\rho^{pn}(rz) ] \nonumber \\ =& i\rho^{np}(rz) + c.c.,\\
\rho_{3}(\bbox{r}) =&\rho^{nn}(rz) -\rho^{pp}(rz) \nonumber \\
                  =& \frac {1} {2} \biggr(\rho^{nn}(rz) -
\rho^{pp}(rz) + c.c. \biggr).
\label{eq:rho3}
\end{align}
The isospin structure of the particle density given by Eqs.~(\ref{eq:rho0})-(\ref{eq:rho3}) is identical
for all the following particle-hole
densities, and these expressions shall not be repeated in the following.
\item[(b)] Kinetic density
\begin{align}
\tau_m (\bbox{r}) = (\bbox{\nabla} \cdot \bbox{\nabla} ')\rho_m(\bbox{r}, \bbox{r}')
\biggr|_{ \bbox{r} = \bbox{r}'} = \sum_{tt'}\tau^{tt'}(rz) \tau_{t't}^{m},
\end{align}
where,
\begin{align}
\tau^{tt'}(rz) =& \sum_{k}\biggr\{
[\partial_{r} V^{+\ast}_{kt}][\partial_{r} V^{+}_{kt'}]
+[\partial_{r} V^{-\ast}_{kt}][ \partial_{r}V^{-}_{kt'}] \nonumber \\
&+ \frac{(\Lambda^{-})^{2}}{r^2} V^{+\ast}_{kt}V^{+}_{kt'}
+ \frac{ (\Lambda^{+})^{2}}{r^2} V^{-\ast}_{kt}V^{-}_{kt'} \nonumber \\
&+ [\partial_{z} V^{+\ast}_{kt}][\partial_{z} V^{+}_{kt'}]
+[\partial_{z} V^{-\ast}_{kt}][ \partial_{z} V^{-}_{kt'}]\biggr\}  .
\end{align}

\item[(c)] Pseudovector spin density
\begin{align}
\bbox{s}_{m}(\bbox{r}) = \bbox{s}_{m}(\bbox{r},\bbox{r}') \biggl|_{\bbox{r}  = \bbox{r}'} = \sum_{tt'} \bbox{s}^{tt'}(rz) \tau_{t't}^{m},
\end{align}
where
\begin{align}
\bbox{s}^{tt'}(rz)=&
\sum_k [
\bbox{e}_r s_r^{tt'}(rz) + \bbox{e}_\phi s_\phi^{tt'}(rz) + \bbox{e}_z s_z^{tt'}(rz)
] \nonumber \\
=&
\sum_{k}\biggr\{ \bbox{e}_{r}[V^{-\ast}_{kt} V^{+}_{kt'}
+V^{+\ast}_{kt} V^{-}_{kt'}]\nonumber\\
&+ \bbox{e}_{\phi}i[V^{+\ast}_{kt} V^{-}_{kt'}-V^{-\ast}_{kt} V^{+}_{kt'}]\nonumber\\
&+ \bbox{e}_{z}[ V^{+\ast}_{kt} V^{+}_{kt'} -V^{-\ast}_{kt} V^{-}_{kt'}]\biggr\}.
\end{align}

\item[(d)] Pseudovector spin-kinetic density
\begin{align}
\bbox{T}_m(\bbox{r}) =& \biggr[(\bbox{\nabla} \cdot \bbox{\nabla} ')\bbox{s}_m(\bbox{r}, \bbox{r}') \biggr]_{ \bbox{r} = \bbox{r}'} \nonumber \\
=&\sum_{tt'}\bbox{T}^{tt'}(rz) \tau_{t't}^m,
\end{align}
where
\begin{align}
\bbox{T}^{tt'}(rz) =&  \sum_{k}\biggr\{ \bbox{e}_{r}\biggr( [\partial_{r}V^{+\ast}_{kt}][ \partial_{r} V^{-}_{kt'}]+\frac{\Lambda^{+}\Lambda^{-}}{r^2}V^{+\ast}_{kt}V^{-}_{kt'} \nonumber \\
& +[\partial_{z} V^{+\ast}_{kt}][ \partial_{z} V^{-}_{kt'}]+[\partial_{r} V^{-\ast}_{kt}][ \partial_{r} V^{+}_{kt'}] \nonumber \\
& + \frac{\Lambda^{-}\Lambda^{+}}{r^2}V^{-\ast}_{kt}V^{+}_{kt'}+[\partial_{z} V^{-\ast}_{kt}][ \partial_{z} V^{+}_{kt'}]\biggr)\nonumber\\
& +i\bbox{e}_{\phi}\biggr( [\partial_{r} V^{+\ast}_{kt}][\partial_{r} V^{-}_{kt'}]+\frac{\Lambda^{-}\Lambda^{+}}{r^2}V^{+\ast}_{kt}V^{-}_{kt'} \nonumber \\
& +[\partial_{z} V^{+\ast}_{kt}][ \partial_{z} V^{-}_{kt'}]-[\partial_{r} V^{-\ast}_{kt}][ \partial_{r} V^{+}_{kt'}]\nonumber \\
& - \frac{\Lambda^{-}\Lambda^{+}}{r^2}V^{-\ast}_{kt}V^{+}_{kt'} -[\partial_{z} V^{-\ast}_{kt}][ \partial_{z} V^{+}_{kt'}]\biggr)\nonumber \\
& +\bbox{e}_z\biggr([\partial_{r} V^{+\ast}_{kt}][\partial_{r} V^{+}_{kt'}]+\frac{\Lambda^{-2}}{r^2}V^{+\ast}_{kt}V^{+}_{kt'} \nonumber \\
& +[\partial_{z} V^{+\ast}_{kt}][\partial_{z} V^{+}_{kt}]-[\partial_{r} V^{-\ast}_{kt}][ \partial_{r} V^{-}_{kt'}] \nonumber \\
& - \frac{\Lambda^{+2}}{r^2}V^{-\ast}_{kt}V^{-}_{kt'}-[\partial_{z} V^{-\ast}_{kt}][ \partial_{z}V^{-}_{kt'}]\biggr) \biggr\}.
\end{align}

\item[(e)] Vector current density
\begin{align}
\bbox{j}_{m}(\bbox{r}) =& \frac{1}{2i}\biggr[(\bbox{\nabla}-\bbox{\nabla}')\rho_{m}(\bbox{r}, \bbox{r}')\biggr]_{\bbox{r}=\bbox{r}'}\nonumber \\
=& \sum_{tt'}\bbox{j}^{tt'}(rz) \tau_{t't}^m,
\end{align}
where
\begin{align}
\bbox{j}^{tt'}(rz) =& \frac{1}{2i}\sum_{k}\biggr\{ \bbox{e}_r\biggr([\partial_{r} V^{+\ast}_{kt}]V^{+}_{kt'}+[\partial_{r} V^{-\ast}_{kt}] V^{-}_{kt'}\nonumber\\
&  - V^{+\ast}_{kt}[\partial_{r}V^{+}_{kt'}]-V^{-\ast}_{kt} [\partial_{r}V^{-}_{kt'}]\biggr)\nonumber\\
&-2i\bbox{e}_{\phi}\biggr( \frac{\Lambda^{-}}{r}V^{+\ast}_{kt}V^{+}_{kt'}+ \frac{\Lambda^{+}}{r}V^{-\ast}_{kt} V^{-}_{kt'}\biggr)\nonumber\\
&+\bbox{e}_z\biggr( [\partial_{z}V^{+\ast}_{kt}]V^{+}_{kt'}+[\partial_{z} V^{-\ast}_{kt}] V^{-}_{kt'}\nonumber\\
&  - V^{+\ast}_{kt}[\partial_{z} V^{+}_{kt'}]-V^{-\ast}_{kt} [\partial_{z} V^{-}_{kt'}]\biggr) \biggr\}.
\end{align}
\item[(f)]  Tensor spin current density
\begin{align}
{\sf J}_{m}^{ij}=\frac{1}{2i}(\nabla_i-\nabla'_i)s_m^j(\bbox{r}, \bbox{r}')\biggr|_{\bbox{r}=\bbox{r}'}
=\sum_{tt'}{\sf J}^{tt'}_{ij}(rz) \tau^m_{t't}  .
\end{align}
Explicit expressions of the components are
\begin{align}
{\sf J}_{r\phi}^{tt'}(rz) =& \frac{1}{2}\sum_{k}\biggr\{[\partial_{r} V^{+\ast}_{kt}]V^{-}_{kt'}-V^{+\ast}_{kt}[\partial_{r} V^{-}_{kt'}]\nonumber\\
& -[\partial_{r} V^{-\ast}_{kt}]V^{+}_{kt'}+V^{-\ast}_{kt}[\partial_{r} V^{+}_{kt'}]\biggr\}, \\
{\sf J}_{rz}^{tt'}(rz) =& \frac{1}{2i}\sum_{k}\biggr\{[\partial_{r} V^{+\ast}_{kt}]V^{+}_{kt'}-V^{+\ast}_{kt}[\partial_{r} V^{+}_{kt'}]\nonumber\\
& -[\partial_{r} V^{-\ast}_{kt}]V^{-}_{kt'}+V^{-\ast}_{kt}[\partial_{r} V^{-}_{kt'}]\biggr\}, \\
{\sf J}_{\phi z}^{tt'}(rz) =& -\sum_{k}\biggr\{ \frac{\Lambda^{-}}{r}V^{+\ast}_{kt}V^{+}_{kt'}-\frac{\Lambda^{+}}{r}V^{-\ast}_{kt}V^{-}_{kt'}\biggr\}, \\
{\sf J}_{z\phi}^{tt'}(rz) =& \frac{1}{2}\sum_{k}\biggr\{[\partial_{z} V^{+\ast}_{kt}]V^{-}_{kt'}-V^{+\ast}_{kt}[\partial_{z} V^{-}_{kt'}]\nonumber\\
& -[\partial_{z} V^{-\ast}_{kt}]V^{+}_{kt'}+V^{-\ast}_{kt}[\partial_{z} V^{+}_{kt'}]\biggr\}, \\
{\sf J}_{zr}^{tt'}(rz) =& \frac{1}{2i}\sum_{k}\biggr\{[\partial_{z} V^{+\ast}_{kt}]V^{-}_{kt'}-V^{+\ast}_{kt}[\partial_{z} V^{-}_{kt'}]\nonumber\\
& +[\partial_{z} V^{-\ast}_{kt}]V^{+}_{kt'}-V^{-\ast}_{kt}[ \partial_{z}V^{+}_{kt'}]\biggr\}, \\
{\sf J}_{\phi r}^{tt'}(rz)=&-\frac{1}{2}\sum_{k}\biggr\{ \frac{\Lambda^{-}}{r}V^{+\ast}_{kt}V^{-}_{kt'}+\frac{\Lambda^{-}}{r}V^{-\ast}_{kt}V^{+}_{kt'}\nonumber\\
& +\frac{\Lambda^{+}}{r}V^{+\ast}_{kt}V^{-}_{kt'}+\frac{\Lambda^{+}}{r}V^{-\ast}_{kt}V^{+}_{kt'}\biggr\}, \\
{\sf J}_{zz}^{tt'}(rz) =& \frac{1}{2i}\sum_{k}\biggr\{[\partial_{z} V^{+\ast}_{kt}]V^{+}_{kt'}-V^{+\ast}_{kt}[ \partial_{z}V^{+}_{kt'}]\nonumber\\
& -[\partial_{z} V^{-\ast}_{kt}]V^{-}_{kt'}+V^{-\ast}_{kt}[\partial_{z} V^{-}_{kt'}]\biggr\}, \\
{\sf J}_{\phi \phi}^{tt'}(rz)=&-\frac{1}{2i}\sum_{k}\biggr\{ \frac{\Lambda^{-}}{r}V^{-\ast}_{kt}V^{+}_{kt'}+\frac{\Lambda^{+}}{r}V^{-\ast}_{kt}V^{+}_{kt'}\nonumber\\
&-\frac{\Lambda^{+}}{r}V^{+\ast}_{kt}V^{-}_{kt'}-\frac{\Lambda^{-}}{r}V^{+\ast}_{kt}V^{-}_{kt'}\biggr\}, \\
{\sf J}_{rr}^{tt'}(rz) =& \frac{1}{2i}\sum_{k}\biggr\{[\partial_{r} V^{+\ast}_{kt}]V^{-}_{kt'}-V^{+\ast}_{kt}[\partial_{r} V^{-}_{kt'}]\nonumber\\
& +[\partial_{r} V^{-\ast}_{kt}]V^{+}_{kt'}-V^{-\ast}_{kt}[\partial_{r} V^{+}_{kt'}]\biggr\}.
\end{align}
\end{itemize}
The trace, antisymmetric, and symmetric parts of the tensor spin-current density are given by
\begin{align}
  J_k(\bbox{r}) =& \sum_{a=x,y,z} {\sf J}_{kaa}(\bbox{r}), \\
  \bbox{J}_{ka}(\bbox{r}) =& \sum_{b,c=x,y,z} \bbox{\epsilon}_{abc} {\sf J}_{kbc}(\bbox{r}), \\
  \underline{\sf J}_{kab}(\bbox{r}) =& \frac{1}{2}{\sf J}_{kab}(\bbox{r}) + \frac{1}{2}{\sf J}_{kba}(\bbox{r})
 - \frac{1}{3} J_k(\bbox{r}) \delta_{ab}.
\end{align}

\subsection{Derivatives of the densities}
\begin{itemize}
\item[(a)] Divergence of tensor spin current density
\begin{align}
[\bbox{\nabla} \cdot \bbox{J}_m(\bbox{r})] =& \frac{1}{2i}\sum_{kss'tt'}\biggr\{ \bbox{\nabla} V_k(\bbox{r} s't')\cdot (\bbox{\nabla} \times \hat{\bbox{\sigma}}_{s's})V_k^\ast(\bbox{r} st) \nonumber \\
& -\bbox{\nabla} V_k^\ast(\bbox{r} st)\cdot(\bbox{\nabla} \times \hat{\bbox{\sigma}}_{s's})V_k(\bbox{r} s't')\biggr\}\tau_{t't}^m \nonumber \\
&\hspace*{-1cm}=\sum_{tt'} \biggr( DJ_r^{tt'}(rz) +
DJ_\phi^{tt'}(rz)  + DJ_z^{tt'}(rz) \biggr) \tau_{t't}^{m},
\end{align}
where
\begin{align}
DJ_{r}^{tt'}(rz) =& \frac{1}{2}\sum_{k}\biggr\{-\frac{\Lambda^{-}}{r}[\partial_r V^{+}_{kt'}]V^{+\ast}_{kt}-\frac{\Lambda^{-}}{r}[\partial_r V^{+\ast}_{kt}]V^{+}_{kt'}\nonumber\\
&+[\partial_r V^{+\ast}_{kt}][\partial_z V^{-}_{kt'}]-[\partial_r V^{-\ast}_{kt}][\partial_z V^{+}_{kt'}]\nonumber\\
&-[\partial_r V^{-}_{kt'}][\partial_z V^{+\ast}_{kt}]+[\partial_r V^{+}_{kt'}][\partial_z V^{-\ast}_{kt}]\nonumber\\
& +\frac{\Lambda^{+}}{r}[\partial_r V^{-}_{kt'}]V^{-\ast}_{kt}+\frac{\Lambda^{+}}{r}[\partial_r V^{-\ast}_{kt}]V^{-}_{kt'}\biggr\}, \\
DJ_{\phi}^{tt'}(rz) =& -\frac{1}{2}\sum_{k}\biggr\{\frac{\Lambda^{-}}{r}V^{+}_{kt'}[\partial_r V^{+\ast}_{kt}]+\frac{\Lambda^{-}}{r}V^{+\ast}_{kt}[\partial_r V^{+}_{kt'}]\nonumber\\
&-\frac{\Lambda^{-}}{r}V^{+}_{kt'}[\partial_z V^{-\ast}_{kt}]-\frac{\Lambda^{+}}{r}V^{-\ast}_{kt}[\partial_z V^{+}_{kt'}]\nonumber\\
&-\frac{\Lambda^{+}}{r}V^{-}_{kt'}[\partial_z V^{+\ast}_{kt}]-\frac{\Lambda^{-}}{r}V^{+\ast}_{kt}[\partial_z V^{-}_{kt'}]\nonumber\\
& -\frac{\Lambda^{+}}{r}V^{-}_{kt'}[\partial_r V^{-\ast}_{kt}]-\frac{\Lambda^{+}}{r}V^{-\ast}_{kt}[\partial_r V^{-}_{kt'}]\biggr\}, \\
 DJ_{z}^{tt'}(rz) =& - \frac{1} {2} \sum_{k}\biggr\{[\partial_z V^{+}_{kt'}][\partial_r V^{-\ast}_{kt}]-\frac{\Lambda^{+}}{r}[\partial_z V^{+}_{kt'}]V^{-\ast}_{kt}\nonumber\\
&-[\partial_z V^{-\ast}_{kt}][\partial_r V^{+}_{kt'}]-\frac{\Lambda^{-}}{r}[\partial_z V^{-\ast}_{kt}] V^{+}_{kt'}\nonumber\\
&-[\partial_z V^{-}_{kt'}][\partial_r V^{+\ast}_{kt}]-\frac{\Lambda^{-}}{r}[\partial_z V^{-}_{kt'}]V^{+\ast}_{kt}\nonumber\\
& +[\partial_z V^{+\ast}_{kt}][\partial_r V^{-}_{kt'}]-\frac{\Lambda^{+}}{r}[\partial_z V^{+\ast}_{kt}]V^{-}_{kt'}\biggr\}.
\end{align}
Isospin components are given by
\begin{align}
\bbox{\nabla} \cdot \bbox{J}_0(\bbox{r}) =& \sum_{i=(r,\phi,z)} \biggr( DJ_{i}^{nn}(rz)+ DJ_{i}^{pp}(rz)
\biggr), \\
\bbox{\nabla} \cdot \bbox{J}_1(\bbox{r}) =& \sum_{i=(r,\phi,z)}\biggr( DJ_{i}^{np}(rz)+
DJ_{i}^{pn}(rz)\biggr), \\
\bbox{\nabla} \cdot \bbox{J}_2(\bbox{r}) =& i\sum_{i=(r,\phi,z)}  \biggr(
DJ_{i}^{np}(rz)-DJ_{i}^{pn}(rz)\biggr), \\
\bbox{\nabla} \cdot \bbox{J}_3(\bbox{r}) =& \sum_{i=(r,\phi,z)} \biggr( DJ_{i}^{nn}(rz)- DJ_{i}^{pp}(rz)\biggr).
\end{align}
\item[(b)] Curl of current density
\begin{align}
\bbox{\nabla} \times \bbox{j}_m(\bbox{r}) =\sum_{tt'} (\bbox{\nabla} \times \bbox{j})^{tt'}(rz)~\tau_{t't}^{m},
\end{align}
where
\begin{align}
(\bbox{\nabla} \times \bbox{j})^{tt'}(rz) =& \sum_{k}\biggr\{ \nonumber \\
& \bbox{e}_r\biggr(\frac{\Lambda^{-}}{r}V^{+\ast}_{kt}[\partial_zV^{+}_{kt'}]+ \frac{\Lambda^{+}}{r}V^{-\ast}_{kt}[\partial_z V^{-}_{kt'}]\nonumber\\
&+\frac{\Lambda^{-}}{r}[\partial_z V^{+\ast}_{kt}]V^{+}_{kt'}+ \frac{\Lambda^{+}}{r}[\partial_z V^{-\ast}_{kt}]V^{-}_{kt'}\biggr)\nonumber\\
&+i\bbox{e}_{\phi}\biggr([\partial_z V^{+\ast}_{kt}][\partial_r V^{+}_{kt'}]+ [\partial_z V^{-\ast}_{kt}][\partial_r V^{-}_{kt'}]\nonumber\\
&-[\partial_r V^{+\ast}_{kt}][\partial_z V^{+}_{kt'}]-[\partial_r V^{-\ast}_{kt}][\partial_z V^{-}_{kt'}]\biggr)\nonumber\\
& -\bbox{e}_z\biggr(\frac{\Lambda^{-}}{r}[\partial_r V^{+\ast}_{kt}]V^{+}_{kt'}+ \frac{\Lambda^{+}}{r}[\partial_r V^{-\ast}_{kt}]V^{-}_{kt'}\nonumber\\
&+\frac{\Lambda^{-}}{r}V^{+\ast}_{kt}[\partial_r V^{+}_{kt'}]+ \frac{\Lambda^{+}}{r}V^{-\ast}_{kt}[\partial_r V^{-}_{kt'}]\biggr)\biggr\}
\end{align}
\item[(c)] Curl of spin density
\begin{align}
\bbox{\nabla} \times \bbox{s}_m(\bbox{r}) =\sum_{tt'} (\bbox{\nabla} \times \bbox{s})^{tt'}(rz)~\tau_{t't}^{m},
\end{align}
where
\begin{align}
(\bbox{\nabla} \times \bbox{s})^{tt'}(rz) =& i \sum_{k}\biggr\{ \bbox{e}_r\biggr([\partial_z V^{-\ast}_{kt}]V^{+}_{kt'}+ V^{-\ast}_{kt}[\partial_z V^{+}_{kt'}]\nonumber\\
&-[\partial_z V^{+\ast}_{kt}]V^{-}_{kt'}- V^{+\ast}_{kt}[\partial_z V^{-}_{kt'}]\biggr)\nonumber\\
&-i \bbox{e}_{\phi}\biggr([\partial_z V^{-\ast}_{kt}]V^{+}_{kt'}+ V^{-\ast}_{kt}[\partial_z V^{+}_{kt'}]\nonumber\\
&+[\partial_z V^{+\ast}_{kt}]V^{-}_{kt'}+V^{+\ast}_{kt}[\partial_z V^{-}_{kt'}]\nonumber\\
& -[\partial_r V^{+\ast}_{kt}]V^{+}_{kt'}- V^{+\ast}_{kt}[\partial_r V^{+}_{kt'}]\nonumber\\
&+[\partial_r V^{-\ast}_{kt}]V^{-}_{kt'}+V^{-\ast}_{kt}[\partial_r V^{-}_{kt'}]\biggr)\nonumber\\
&+\bbox{e}_z\biggr([\partial_r V^{+\ast}_{kt}]V^{-}_{kt'}+ V^{+\ast}_{kt}[\partial_r V^{-}_{kt'}]\nonumber\\
&-[\partial_r V^{-\ast}_{kt}]V^{+}_{kt'}- V^{-\ast}_{kt}[\partial_r V^{+}_{kt'}]\biggr)\biggr\}.
\end{align}
\item[(d)] Laplacian of $\rho$
\begin{align}
\nabla^{2}\rho_m(\bbox{r}) = \sum_{tt'} \biggr( 2 \tau^{tt'}(rz) + L^{tt'}(rz)\biggr)~~~\tau_{t't}^{m},
\end{align}
where
\begin{align}
L^{tt'}(rz) =& \sum_{k}\biggr\{ \bar{V}^{+\ast}_{kt} [\nabla^{2} \bar{V}^{+}_{kt'} ] +
\bar{V}^{-\ast}_{kt} [\nabla^{2} \bar{V}^{-}_{kt'}]\nonumber\\
& + \bar{V}^{+}_{kt'} [\nabla^{2}\bar{V}^{+\ast}_{kt} ]
+\bar{V}^{-}_{kt'} [\nabla^{2} \bar{V}^{-\ast}_{kt} ]\biggr\}, \\
\bar{V}^{+}_{kt} =& V^{+}_{kt} e^{i \Lambda^{-} \phi}, \\
\bar{V}^{-}_{kt} =& V^{-}_{kt} e^{i \Lambda^{+} \phi}.
\end{align}

\item[(e)] Laplacian of $\bbox{s}$
\begin{align}
\nabla^{2}\bbox{s}_m(\bbox{r}) =& \sum_{tt'} \biggr( 2 \bbox{T}^{tt'}(rz) + \bbox{S}^{tt'}(rz) \nonumber \\
&- \frac {{\bbox{e}}_r} { r^2 } s_r^{tt'}(rz) - \frac {{\bbox{e}}_\phi} {r^2}  s_\phi^{tt'} (rz) \biggr)\tau_{t't}^{m},
\end{align}
where
\begin{align}
\bbox{S}^{tt'}(rz) =&  \sum_{k}\biggr\{ {\bbox{e}}_r \biggr( [\nabla^{2} \bar{V}^{-\ast}_{kt}] \bar{V}^{+}_{kt'} +\bar{V}^{-\ast}_{kt} [\nabla^{2} \bar{V}^{+}_{kt'}]\nonumber\\
&+[\nabla^{2} \bar{V}^{+\ast}_{kt} ] \bar{V}^{-}_{kt'} + \bar{V}^{+\ast}_{kt} [\nabla^{2} \bar{V}^{-}_{kt'} ]\biggr)\nonumber\\
&+ i {\bbox{e}}_\phi \biggr( -[\nabla^{2} \bar{V}^{-\ast}_{kt}] \bar{V}^{+}_{kt'} - \bar{V}^{-\ast}_{kt} [\nabla^{2} \bar{V}^{+}_{kt'} ]\nonumber\\
&+[\nabla^{2} \bar{V}^{+\ast}_{kt}] \bar{V}^{-}_{kt'} + \bar{V}^{+\ast}_{kt} [\nabla^{2} \bar{V}^{-}_{kt'} ]\biggr)\nonumber\\
&+ {\bbox{e}}_z \biggr( [\nabla^{2} \bar{V}^{+\ast}_{kt} ] \bar{V}^{+}_{kt'} + \bar{V}^{+\ast}_{kt} [\nabla^{2} \bar{V}^{+}_{kt'} ]\nonumber\\
&-[\nabla^{2} \bar{V}^{-\ast}_{kt} ]\bar{V}^{-}_{kt'}- \bar{V}^{-\ast}_{kt} [\nabla^{2} \bar{V}^{-}_{kt'}]\biggr) \biggr\}.
\end{align}
\end{itemize}


%

\end{document}